\documentstyle[aps,preprint]{revtex}
\begin{document}

\newcommand{\pp}[1]{\phantom{#1}}
\newcommand{\be}{\begin{eqnarray}}
\newcommand{\ee}{\end{eqnarray}}
\newcommand{\ve}{\varepsilon}
\newcommand{\Tr}{{\rm Tr\,}}
\newtheorem{th}{Theorem}
\newtheorem{lem}[th]{Lemma}


\title{
Nambu-Type Generalization of the Dirac Equation
}
\author{Marek Czachor}
\address{
Wydzia{\l}  Fizyki Technicznej i Matematyki Stosowanej\\
 Politechnika Gda\'{n}ska,
ul. Narutowicza 11/12, 80-952 Gda\'{n}sk, Poland\\
email: mczachor@sunrise.pg.gda.pl
}
\maketitle
\begin{abstract}
Nonlinear generalization of the Dirac equation extending
the standard paradigm of 
nonlinear Hamiltonians is discussed. 
``Faster-than-light telegraphs" are
absent for all theories formulated within the new framework. 
A new metric for
infinite dimensional Lie algebras
associated with Lie-Poisson dynamics is introduced.
\end{abstract}

\section{Introduction}

In linear quantum mechanics an algebra of observables is
associative and therefore can be naturally related to random
variables measured in experiments. In nonlinear quantum
mechanics observables (at least Hamiltonians) are represented by
nonlinear operators which 
do not naturally lead to a notion of eigenvalue.  
The problem is that although it is quite
easy to define various nonlinear eigenvalues \cite{BBM1,W2,MR}, 
it is difficult
to consistently
associate with them values of random variables measured in 
experiments (cf. the discussion of propositions in \cite{Jor}
and problems with probability interpretation discussed in
\cite{pra,doktorat}). There are several possible ways out of the
difficulty. First, one may hope that a consistent theory of
measurement will be formulated also for fields evolving
nonlinearly. An attempt of Weinberg \cite{W2} goes in this
direction, but the theory he proposed was based on several
arbitrary elements including the (nonrelativistic in nature)
assumption that a nonlinear theory may involve nonlinear
Hamiltonians and linear momenta. Second, in the
 approach of Jordan
\cite{Jor1} (similar ideas can
be found in \cite{doktorat}) 
one distinguishes between observables and
generators of symmetry transformations (say, Hamiltonian and
energy). Unfortunately, in 
practical applications the idea turns out to possess
ambiguities as well (some of them are mentioned in \cite{pra} 
in the context of two-level atoms).
The third direction was initiated by Mielnik \cite{Mielnik} 
who proposed to
discuss the probability interpretation of generalized theories
at the abstract level of convex figures of states. A practical
difficulty is that it is very difficult to apply such ideas in
concrete situations since we have to know the ``shape" of the
figure of mixed states. 
The fourth obvious possibility would be to find a nonlinear quantum
mechanics where Hamiltonians and other observables could be
simply kept linear. 

A general framework which on one hand allows 
for such generalizations, and on the other contains the
Weinberg-type theory as a particular case, is
presented in this Letter. We start with rewriting the relativistic 
density matrix formalism of the Dirac equation in a form of a
Nambu-type bracket introduced by Bia{\l}ynicki-Birula and
Morrison \cite{Mor}, but here generalized to relativistic and
multi-particle systems. Next, we prove that a large class of
Nambu-type theories based on the triple bracket is free of the
causality problems discussed in the context of the Weinberg
theory \cite{Gisin1,Gisin2,Polchinski,Jordan,MC1,MC2}. 
We prove also a theorem stating that at least for a large class
of initial conditions a solution of the generalized equation is
a density matrix. These theorems generalize earlier results
proved by Polchinski and Jordan for Weinberg's nonlinear quantum
mechanics. 

The paper is organized as follows. In Sec.~\ref{s1} we describe
the Hamiltonian formulation of the Dirac equation. Elements of
this formulation can be found in literature but 
in a form which is not very helpful
from the point of view of nonlinear Nambu-type generalizations
discussed in Sec.~\ref{s}.
Sec.~\ref{s2} presents a compact abstract index formalism which
exploits formal analogies between density matrices and
world-vectors. As a by-product we discuss an interesting metric
structure associated with infinite dimensional Lie algebras
which can be used in situations where the standard
Killing-Cartan finite-dimensional metric does not exist. In
Sec.~\ref{s3} we prove two important technical lemmas and the theorem
on nonexistence of ``faster-than-light telegraphs". In Sec.~\ref{s4}
we discuss the density matrix interpretation of solutions of the
Nambu-type solutions of the generalized Dirac equation. Finally,
in Sec.~\ref{s5} we discuss examples showing that even keeping
Hamiltonians linear one can obtain a nonlinear dynamics. 

In a separate paper we shall present an attempt of a physical
interpretation of the generators of the Nambu-type dynamics. 
The interpretation relates the second generator to R\'enyi's
$\alpha$-entropies \cite{Renyi} and leads to interesting 
correlations between linearity of evolution and possibilities
of gaining information by a quantum system.

\section{Hamiltonian formulation of the Dirac equation}
\label{s1}

The Dirac equation in the Minkowski space
representation is ($\hbar=1$)
\begin{eqnarray}
i\nabla_{AA'}\psi^A&=&\frac{m}{\sqrt{2}} \xi_{A'},\label{D1}\\
i\nabla^{AA'}\xi_{A'}&=&\frac{m}{\sqrt{2}} \psi^{A}\label{D2}
\end{eqnarray}
where $\nabla_{AA'}=\nabla^ag_{aAA'}$ and
 $g_{aAA'}$ denote Infeld-van
der Waerden tensors \cite{PR} (all indices should be understood
in the abstract sense of Penrose). To simplify formulas we shall
assume that the derivative $\nabla^a$ does not contain
four-potentials, but the Lie-Poisson and Lie-Nambu algebraic
structures we shall derive below would not be changed if we had
considered Dirac fields coupled to the Maxwell field
(c.f. \cite{doktorat}). 

Contracting Eq.~(\ref{D1}) with $g{_a}^{BA'}$, Eq.~(\ref{D2}) 
with $g_{aAB'}$, using identities (\ref{id1}), 
(\ref{id2}) from the Appendix and the (anti-)self-duality
properties of the generators  we obtain 
\begin{eqnarray}
P_{a}\psi^A&=&2\,{P^b}{^*
\sigma}_{ba}{^{A}}{_{B}}\psi^B +
\sqrt{2}m g{_a}^{AB'}\xi_{B'},\label{D1'}\\
P_{a}\xi_{A'}&=&2\,{P^b}{^*\bar 
\sigma}_{ba}{_{A'}}{^{B'}}\xi_{B'} +
\sqrt{2}m g_{aBA'}
\psi^{B},\label{D2'}
\end{eqnarray}
where $P_a=i\nabla_a$. This form of the Dirac equation can be
found in \cite{Kalnay} and in analogous form in \cite{Hammer}
although none of those authors used an explicitly spinor
formulation. Let $n^a$ be an arbitrary
future-pointing and normalized ($n^an_a=1$) timelike
world-vector. Denote $n^aP_a=in^a\nabla_a=in\cdot \nabla$.
Contracting Eqs.~(\ref{D1'}), (\ref{D2'}) with $n^a$, then switching
from world-vector to spinor indices, using (\ref{gs1}), 
(\ref{gs2}) and $n_{AA'}n^{BA'}=\ve{_A}{^B}/2$, we obtain 
\begin{eqnarray}
i\,n\cdot\nabla\psi^A&=&i\,n^{BB'}\nabla{^{A}}{_{B'}}\psi_B +
\frac{m}{\sqrt{2}}n^{AB'}\xi_{B'},\label{D1''}\\
i\,n\cdot\nabla\xi_{A'}&=&i\,n^{BB'}\nabla{_{BA'}}
\xi_{B'} +
\frac{m}{\sqrt{2}}n_{BA'}
\psi^{B},\label{D2''}
\end{eqnarray}
where $n^{AB'}=n^ag_a{^{AB'}}$. The complex-conjugated equations
are
\begin{eqnarray}
-i\,n\cdot\nabla\bar \psi^{A'}
&=&-i\,n^{BB'}\nabla{_{B}}{^{A'}}\bar \psi_{B'} +
\frac{m}{\sqrt{2}}n^{BA'}\bar \xi_{B},\label{ccD1''}\\
-i\,n\cdot\nabla\bar \xi_{A}&=&-i\,n^{BB'}\nabla{_{AB'}}
\bar \xi_{B} +
\frac{m}{\sqrt{2}}n_{AB'}
\bar \psi^{B'}.\label{ccD2''}
\end{eqnarray}
Let $d\mu(n,x)=e_{abcd}n^a\,dx^b\wedge dx^c\wedge dx^d$ be the
measure on the spacelike hyperplane $x^an_a -\tau=0$. Define 
the norm 
\be
\parallel \Psi\parallel^2 =
\int d\mu(n,x) n^{AA'}\Bigl(
\psi_{A}\bar \psi_{A'}+
\bar \xi_{A}\xi_{A'}
\Bigr)\label{norm}
\ee
where 
\be
\Psi_\alpha=
\left(
\begin{array}{c}
\psi_A\\
\xi_{A'}
\end{array}
\right).
\ee
The Hamilton equations equivalent to (\ref{D1''}), (\ref{D2''}),
(\ref{ccD1''}), and (\ref{ccD2''}) can be obtained from the
Hamiltonian function
\be
H &=& H[\Psi,\bar \Psi]=H[\psi,\bar \psi,\xi,\bar \xi]\\
&=&\int
d\mu(n,x)
(\bar \psi_{A'},\bar \xi_{A})
\left(
\begin{array}{cc}
n^{AA'} & 0\\
0 & n^{AA'}
\end{array}
\right)
\left(
\begin{array}{cc}
i\overrightarrow{\nabla}_{AB'} & -\frac{m}{\sqrt{2}}\ve_{AB}\\
\frac{m}{\sqrt{2}}\ve_{A'B'} & i\overrightarrow{\nabla}_{BA'}
\end{array}
\right)
\left(
\begin{array}{cc}
n^{BB'} & 0\\
0 & n^{BB'}
\end{array}
\right)
\left(
\begin{array}{c}
\psi_B\\
\xi_{B'}
\end{array}
\right)\nonumber\\
&=&\int
d\mu(n,x)
(\bar \psi_{A'},\bar \xi_{A})
\left(
\begin{array}{cc}
n^{AA'} & 0\\
0 & n^{AA'}
\end{array}
\right)
\left(
\begin{array}{cc}
-i\overleftarrow{\nabla}_{AB'} & -\frac{m}{\sqrt{2}}\ve_{AB}\\
\frac{m}{\sqrt{2}}\ve_{A'B'} & -i\overleftarrow{\nabla}_{BA'}
\end{array}
\right)
\left(
\begin{array}{cc}
n^{BB'} & 0\\
0 & n^{BB'}
\end{array}
\right)
\left(
\begin{array}{c}
\psi_B\\
\xi_{B'}
\end{array}
\right),\nonumber
\ee
where $\overleftarrow{\nabla}$ and $\overrightarrow{\nabla}$ 
act to the left and to the right, respectively. 
Let $I_{AA'}=2n_{AA'}$ and $\omega^{AA'}=n^{AA'}$, and let us
denote the directional derivative $n\cdot \nabla$ by a dot.
The Dirac equation can be written as classical Hamilton
equations. 
\be
\begin{array}{rr}
{i\dot \psi_A=I_{AA'}\frac{\delta H}{\delta \bar \psi_{A'}}}, &
{i\dot \xi_{A'}=I_{AA'}\frac{\delta H}{\delta \bar \xi_{A}}},\\
{-i\dot {\bar \psi}_{A'}=I_{AA'}\frac{\delta H}{\delta \psi_{A}}},
&
{-i\dot {\bar \xi}_{A}=I_{AA'}\frac{\delta H}{\delta \xi_{A'}}},
\end{array}
\ee
or
\be
\begin{array}{rr}
{i\omega^{AA'}\dot \psi_A=\frac{\delta H}{\delta \bar
\psi_{A'}}}, &
{i\omega^{AA'}\dot \xi_{A'}=\frac{\delta H}{\delta \bar \xi_{A}}},\\
{-i\omega^{AA'}\dot {\bar \psi}_{A'}=\frac{\delta H}{\delta
\psi_{A}}}, &
{-i\omega^{AA'}\dot {\bar \xi}_{A}=\frac{\delta H}{\delta \xi_{A'}}}.
\end{array}
\ee

\section{Compact index formulation}
\label{s2}

In what follows we will need an efficient and compact abstract
index formulation of the Hamilton and Poisson equations. 
To begin with let us denote the continuous 
spacetime variables ``$x$" appearing
in the spinor fields
$\psi_A=\psi_A(x)$ and $\xi_{A'}=\xi_{A'}(x)$ by lowercase
boldface Roman indices: $\psi_A=\psi_A(\bbox a)$, 
$\psi_B=\psi_B(\bbox b)$, etc. Second, let us extend the
standard Einstein summation convention by assuming that we
both sum over repeated spinor indices and 
integrate with respect to repeated continuous variables (this
will simplify formulas by eliminating integrals in the same way
the standard convention eliminates sums). Third, the use of
Roman letters for both spinor and spacetime indices alows us to
finally avoid writing  the continuous indices in
formulas explicitly. For example, let $\delta(\bbox a,\bbox a')$ denote the
delta function on the spacelike hyperplane. Armed with the new
convention we can redefine the Poisson tensor and the symplectic
form used in the previous section as follows:
\be
I_{AA'}&=&I_{AA'}(\bbox a,\bbox a')=
2n_{AA'}\delta(\bbox a,\bbox a')\\
\omega^{AA'}&=&\omega^{AA'}(\bbox a,\bbox a')=
n^{AA'}\delta(\bbox a,\bbox a'),
\ee
and the norm (\ref{norm}) becomes
\be
\parallel \Psi\parallel^2 =
\omega^{AA'}\Bigl(
\psi_{A}\bar \psi_{A'}+
\bar \xi_{A}\xi_{A'}
\Bigr).\label{norm'}
\ee
Next define bispinor symplectic form and bispinor Poisson tensor
by
\be
\omega^{\alpha\beta'}&=&
\left(
\begin{array}{cc}
n^{AB'}\delta(\bbox a,\bbox b') & 0\\
0 & n^{BA'}\delta(\bbox b,\bbox a')
\end{array}
\right)=
\left(
\begin{array}{cc}
\omega^{AB'} & 0\\
0 & \omega^{BA'}
\end{array}
\right)\\
I_{\alpha\beta'}&=&
\left(
\begin{array}{cc}
2n_{AB'}\delta(\bbox a,\bbox b') & 0\\
0 & 2n_{BA'}\delta(\bbox b,\bbox a')
\end{array}
\right)=
\left(
\begin{array}{cc}
I_{AB'} & 0\\
0 & I_{BA'}
\end{array}
\right)
\ee
The use of primed bispinor indices is consistent with 
\be
\begin{array}{ll}
{\Psi_\alpha=
\left(
\begin{array}{c}
\psi_A\\
\xi_{A'}
\end{array}
\right),} & 
{\bar \Psi_{\alpha'}=
\left(
\begin{array}{c}
\bar \psi_{A'}\\
\bar \xi_{A}
\end{array}
\right),}
\end{array}.
\ee
Now the norm, the Hamiltonian function, and the equations become
\be
\parallel \Psi\parallel^2 &=&
\omega^{\alpha\alpha'}\Psi_\alpha\bar \Psi_{\alpha'},\\
H[\Psi,\bar \Psi]&=&
\omega^{\beta\alpha'}
\bar \Psi_{\alpha'} {H}
_{\beta\gamma'}\Psi_\delta\omega^{\delta\gamma'},\\
i{\dot \Psi}_\beta &=& {H}
_{\beta\gamma'}\Psi_\delta\omega^{\delta\gamma'},\\
-i{\dot {\bar \Psi}}_{\gamma'} &=& 
\omega^{\beta\alpha'} \bar \Psi_{\alpha'}{H}
_{\beta\gamma'}\\
i{\dot \Psi}_\alpha &=& I_{\alpha\alpha'}
\frac{\delta H}{\delta\bar \Psi_{\alpha'}},\\
-i{\dot {\bar \Psi}}_{\alpha'} &=& I_{\alpha\alpha'}
\frac{\delta H}{\delta \Psi_{\alpha}}\\
i\omega^{\alpha\alpha'}{\dot \Psi}_\alpha &=& 
\frac{\delta H}{\delta\bar \Psi_{\alpha'}},\\
-i\omega^{\alpha\alpha'}{\dot {\bar \Psi}}_{\alpha'} &=& 
\frac{\delta H}{\delta \Psi_{\alpha}}.
\ee
The next important step follows from the fact that the
Hamiltonian function expressed with the help of the pure-state
density matrix $\rho_{\alpha \alpha'}=\Psi_{\alpha}
\bar \Psi_{\alpha'}$ as 
\be
H[\rho]=\omega^{\alpha\beta'}\omega^{\beta\alpha'}
H_{\alpha\alpha'}\rho_{\beta \beta'}\label{H}
\ee
suggests the identification of pairs of primed and unprimed
bispinor indices with single lowercase italic Roman indices:
$\alpha\alpha'=a$, $\beta\beta'=b$ etc. Introduce now two metric
tensors 
\be
g^{ab}&=&\omega^{\alpha\beta'}\omega^{\beta\alpha'}\label{m1}\\
g_{ab}&=&I_{\alpha\beta'}I_{\beta\alpha'}\label{m2}
\ee
satisfying 
$g^{ab}=g^{ba}$, $g_{ab}=g_{ba}$, $g_{ab}g^{bc}=\delta{_a}{^c}$,  
where $\delta{_a}{^b}=\delta{_{\alpha}}{^{\beta}}
\delta{_{\alpha'}}{^{\beta'}}$
and
\be
\begin{array}{ll}
\delta{_{\alpha}}{^{\beta}}=
\left(
\begin{array}{cc}
\ve{_{A}}{^{B}}\delta(\bbox a,\bbox b) & 0\\
0 & \ve{_{A'}}{^{B'}}\delta(\bbox a',\bbox b')
\end{array}
\right),
&
\delta{_{\alpha'}}{^{\beta'}}=
\left(
\begin{array}{cc}
\ve{_{A'}}{^{B'}}\delta(\bbox a',\bbox b') & 0\\
0 & \ve{_{A}}{^{B}}\delta(\bbox a,\bbox b)
\end{array}
\right).
\end{array}
\ee
The Hamiltonian function (\ref{H}) can be rewritten with the
help of the metric tensor as
\be
H[\rho]=g^{ab}H_a\rho_b.
\ee
The metric will be shown below to be a natural substitute for
the Cartan-Killing metric used in finite-dimensional Lie
algebras and is sometimes used implicitly in quantum optics
\cite{Orl}. Useful are also higher-order tensors which can be
constructed from the symplectic form and the Poisson tensor.
Define 
\be
g^{a_1\dots a_n}&=&\omega^{\alpha_1\alpha'_n}
\omega^{\alpha_2\alpha'_1}\omega^{\alpha_3\alpha'_2}\dots
\omega^{\alpha_{n-1}\alpha'_{n-2}}
\omega^{\alpha_n\alpha'_{n-1}},\label{m1'}\\
g_{a_1\dots a_n}&=&I_{\alpha_1\alpha'_n}
I_{\alpha_2\alpha'_1}I_{\alpha_3\alpha'_2}\dots
I_{\alpha_{n-1}\alpha'_{n-2}}
I_{\alpha_n\alpha'_{n-1}}.\label{m2'}
\ee
Denote $g^{a_1}=\omega^{a_1}$, $g_{a_1}=I_{a_1}$. 
Then 
\be
\Tr \bigl(\rho^n\bigr)=
g^{a_1\dots a_n}\rho_{a_1}\dots \rho_{a_n}=: C_n[\rho].\label{C_n}
\ee
To extend the formalism to the case of $N$ free electrons
consider an $N$-particle, totally antisymmetric state vector
$\Psi^{N}_\alpha=\Psi_{\alpha_1\dots \alpha_N}$. 
The corresponding $N$-particle
Hamiltonian function is
\be
H^{N}&=&
\bar \Psi_{\alpha'_1\dots \alpha'_N}
\omega^{\beta_1\alpha'_1}
\dots
\omega^{\beta_N\alpha'_N}
\Bigl(
H_{\beta_1\gamma'_1}I_{\beta_2\gamma'_2}\dots
I_{\beta_N\gamma'_N}
+
\dots\nonumber\\
&\pp =&
\pp 
{
\bar \Psi_{\alpha'_1\dots \alpha'_N}
\omega^{\beta_1\alpha'_1}
\dots
\omega^{\beta_N\alpha'_N}
\Bigl(
}
\dots +
I_{\beta_1\gamma'_1}\dots I_{\beta_{N-1}\gamma'_{N-1}}
H_{\beta_N\gamma'_N}
\Bigr)
\omega^{\delta_1\gamma'_1}
\dots
\omega^{\delta_N\gamma'_N}
\Psi_{\delta_1\dots \delta_N}.
\ee
Define
\be
\omega^{N{\alpha\alpha'}}&=&
\omega^{\alpha_1\alpha'_1
\dots
\alpha_N\alpha'_N}=
\omega^{\alpha_1\alpha'_1}
\dots
\omega^{\alpha_N\alpha'_N},\\
I^{N}{_{\alpha\alpha'}}&=&
I_{\alpha_1\alpha'_1
\dots
\alpha_N\alpha'_N}=
I_{\alpha_1\alpha'_1}
\dots
I_{\alpha_N\alpha'_N}.
\ee
The Hamilton equations equivalent to the $N$-particle Dirac
equation are
\be
i\dot \Psi{^{N}}_\alpha &=& I{^{N}}_{\alpha\alpha'}
\frac{\delta H^{N}}{\delta\bar \Psi{^{N}}_{\alpha'}},\\
-i\dot {\bar \Psi}{^{N}}_{\alpha'} &=& I{^{N}}
_{\alpha\alpha'}
\frac{\delta H^{N}}{\delta \Psi{^{N}}_{\alpha}}\\
i\omega{^{N}}^{\alpha\alpha'}{\dot \Psi{^{N}}}_\alpha &=& 
\frac{\delta H^{N}}{\delta\bar \Psi{^{N}}_{\alpha'}},\\
-i\omega{^{N}}
^{\alpha\alpha'}{\dot {\bar \Psi}{^{N}}}_{\alpha'} &=& 
\frac{\delta H^{N}}{\delta \Psi{^{N}}_{\alpha}}.
\ee
The remaining definitions are obtained by substitutions
$\omega^{\alpha\alpha'}\to \omega{^{N}}^{\alpha\alpha'}$, 
$I_{\alpha\alpha'}\to I{^{N}}_{\alpha\alpha'}$
in (\ref{m1}), (\ref{m2}), (\ref{m1'}), (\ref{m2'}), 
so that
\be
g{^{N}}^{ab}=g^{a_1b_1}\dots g^{a_Nb_N},\\
g{^{N}}_{ab}=g_{a_1b_1}\dots g_{a_Nb_N}.
\ee

\section{Lie-Poisson and Lie-Nambu brackets associated with the
Dirac equation}
\label{s}

Let $F=F[\Psi,\bar \Psi]=F[\rho]$. The Hamilton equations imply
the Poisson bracket equations
\be
i\, \dot F&=&I_{\alpha\alpha'}
\Bigl(
\frac{\delta F}{\delta \Psi_{\alpha}}
\frac{\delta H}{\delta\bar \Psi_{\alpha'}}
-
\frac{\delta H}{\delta \Psi_{\alpha}}
\frac{\delta F}{\delta\bar \Psi_{\alpha'}}
\Bigr)\label{weinberg}\\
&=&
I_{\alpha\beta'}\rho_{\beta\alpha'}
\Bigl(
\frac{\delta F}{\delta \rho_{\alpha\alpha'}}
\frac{\delta H}{\delta \rho_{\beta\beta'}}
-
\frac{\delta H}{\delta \rho_{\alpha\alpha'}}
\frac{\delta F}{\delta \rho_{\beta\beta'}}
\Bigr).\label{jordan}
\ee
Analogous forms of the Poisson bracket were used in the context of
nonrelativistic nonlinear quantum mechanics by Weinberg
\cite{W2} (RHS of Eq.
(\ref{weinberg})) and Jordan \cite{Jordan} 
(RHS of Eq.~(\ref{jordan})). 
The advantage of (\ref{jordan}) lies in a possibility of using
it for general density matrices (mixed states). 
The RHS of (\ref{jordan}) can be
rewritten in a form of a Lie-Poisson bracket 
\be
\{F,H\}=\rho_a\Omega^a{_{bc}}
\frac{\delta F}{\delta \rho_{a}}
\frac{\delta H}{\delta \rho_{b}},\label{LP}
\ee
where
\be
\Omega^a{_{bc}}=
\delta_{\beta'}{^{\alpha'}}
\delta_{\gamma}{^{\alpha}}
I_{\beta\gamma'}
-
\delta_{\gamma'}{^{\alpha'}}
\delta_{\beta}{^{\alpha}}
I_{\gamma\beta'}\label{O}
\ee
satisfy
\begin{eqnarray}
\Omega^{a}_{{\ }cb}=
-\Omega^{a}_{{\ }bc},
\label{antisymmetry}\\
\Omega^{a}_{{\ }bc}
\Omega^{c}_{{\ }de}
+\Omega^{a}_{{\ }ec}
\Omega^{c}_{{\ }bd}
+\Omega^{a}_{{\ }dc}
\Omega^{c}_{{\ }eb}=0.
\label{structure2}
\end{eqnarray}
and hence are structure constants of an infinite dimensional Lie
algebra. For computational reasons 
it is important to be able to raise and lower indices in the
structure constants.  The standard
Cartan-Killing metric \cite{BR} cannot be used in this context
because of the infinite dimension of the algebra: 
$
\Omega^{c}_{{\ }ad}
\Omega^{d}_{{\
}bc}
$
contains expressions such as
$\delta(0)$ which are not distributions in the Schwartz sense and
such a metric cannot be invertible. The correct metric is given
by (\ref{m1}), (\ref{m2}) \cite{zak}. We find
\be
\Omega{_{abc}}&=&g_{ad}\Omega^d{_{bc}}=
I_{\alpha\beta'}
I_{\beta\gamma'}
I_{\gamma\alpha'}
-
I_{\alpha\gamma'}
I_{\beta\alpha'}
I_{\gamma\beta'}\label{O_}\\
\Omega{^{abc}}&=&g^{bd}g^{ce}\Omega^a{_{de}}=
-\omega^{\alpha\beta'}
\omega^{\beta\gamma'}
\omega^{\gamma\alpha'}
+
\omega^{\alpha\gamma'}
\omega^{\beta\alpha'}
\omega^{\gamma\beta'}\label{O^}
\ee
The $N$-particle generalization is obtained by substituting
$\omega^{\alpha\alpha'}\to \omega{^{N}}^{\alpha\alpha'}$, 
$I_{\alpha\alpha'}\to I{^{N}}_{\alpha\alpha'}$
in (\ref{O^}), (\ref{O_}). 
The form (\ref{O_}) can be used to define the Lie-Nambu bracket
\be
[F,G,H]
=
\Omega{_{abc}}
\frac{\delta F}{\delta \rho_{a}}
\frac{\delta G}{\delta \rho_{b}}
\frac{\delta H}{\delta \rho_{c}},\label{LN}
\ee
and its $N$-particle generalization
\be
[F{^{N}},G{^{N}},H{^{N}}]{^{N}}
=
\Omega{^{N}}{_{abc}}
\frac{\delta F{^{N}}}{\delta \rho{^{N}}_{a}}
\frac{\delta G{^{N}}}{\delta \rho{^{N}}_{b}}
\frac{\delta H{^{N}}}{\delta \rho{^{N}}_{c}}.\label{N-LN}
\ee
An analogous generalized Nambu bracket was discussed in the
nonrelativistic context by Bia{\l}ynicki-Birula and Morrison
\cite{Mor}. The linear Liouville-von Neumann equation 
\be
i\dot \rho_a=\{\rho_a,H\},\label{L-vN}
\ee
where $H=H[\rho]=g^{ab}H_a\rho_b$, can be written as
\be
i\dot \rho_a=[\rho_a,H,S]\label{L-vN'}
\ee
where the ``entropy" 
$S=S[\rho]=g^{ab}\rho_a\rho_b/2=C_2[\rho]/2$ (cf.
Eq.~(\ref{C_n})). The total antisymmetry of the structure
constants implies that $S$ itself commutes with any observable and
hence is a Casimir invariant. 
This is analogous to the original bracket introdued by Nambu
\cite{Nambu} where the structure constants corresponded to the
$o(3)$ Lie algebra and the second generator of evolution was the
squared angular momentum which is also a Casimir invariant of
$o(3)$. The Jordan-Weinberg-type equations discussed
in \cite{Jordan} are of the form (\ref{L-vN}) 
but involve Hamiltonian
functions which can be nonlinear functionals of $\rho$. The
triple bracket equation (\ref{L-vN'}) suggests two additional
possibilities of generalizations of the linear equation 
(\ref{L-vN}): (a) equations where both $H$ and $S$ are
generalized and (b) equations where $H$ is kept linear
but $S\neq C_2/2$. Both possibilities are
interesting and provide a general nonlinear framework for
quantum mechanics which is
beyond the standard paradigm of nonlinear
Schr\"odinger equations. The second possibility is interesting
as the first general scheme allowing for nonlinear extensions 
of quantum mechanics which keeps the algebra of observables
unchanged (all approaches known to the author of this Letter
involve at least nonlinear Hamiltonians). 

\section{General properties of the Lie-Nambu bracket}
\label{s3}

In this section we shall derive two important properties of the
Lie-Nambu bracket which hold independently of the form of $H$
and $S$. Consider an $N$-particle density matrix 
$\rho{^N}_{a}=\rho_{a_1\dots a_N}$. A $K$-particle subsystem
($K\leq N$) is described by observables of the form
\be
F^K&=&g^{Nab}F_{a_1\dots a_K}I_{a_{K+1}\dots a_N}
\rho_{b_1\dots b_N}=
g^{Kab}F_{a_1\dots a_K}
\rho_{b_1\dots b_K},
\ee
where 
\be
\rho_{b_1\dots b_K}&=&
g^{a_{K+1}b_{K+1}}\dots g^{a_{K}b_{N}}
I_{a_{K+1}}\dots I_{a_N}
\rho_{b_1\dots b_Kb_{K+1}\dots b_N}\nonumber\\
&=&
\omega^{N-K\,b_{K+1}\dots b_{N}}
\rho_{b_1\dots b_Kb_{K+1}\dots b_N}
\ee
is the subsystem's reduced density matrix. Consider now two,
$M$- and $(N-M-K)$-particle, 
subsystems which do not overlap (i.e. no particle belongs to
both of them). If their reduced density matrices are 
\be
\rho{^I}_{d}&=&\rho{^I}_{d_1\dots d_M}=
\rho_{d_1\dots d_Md_{M+1}\dots d_N}
\omega^{d_{M+1}\dots d_N},\\
\rho{^{II}}_{e}&=&\rho{^{II}}_{e_{M+K+1}\dots e_N}=
\omega^{e_1\dots e_{M+K}}
\rho_{e_1\dots e_{M+K}e_{M+K+1}\dots e_N}
\ee
then 

\medskip\noindent
{\bf Lemma 1.} 
\be
[\rho{^{I}}_{d},\rho{^{II}}_{e},\,\cdot\,]^N=0.
\ee
{\it Proof\/}: 
Reduced density matrices satisfy
\be
\frac{\delta \rho{^I}_{d}}{\delta \rho_{a_1\dots a_N}}=
\delta^{a_1}_{d_1}\dots \delta^{a_M}_{d_M}
\omega^{a_{M+1}\dots a_N},\quad\quad
\frac{\delta \rho{^{II}}_{e}}{\delta \rho_{b_1\dots b_N}}=
\omega^{b_{1}\dots b_{M+K}}
\delta^{b_{M+K+1}}_{e_{M+K+1}}\dots \delta^{b_N}_{e_N}.
\nonumber
\ee
Consider the expression
\be
X_c
&=&
\Omega{^{N}}{_{abc}}
\frac{\delta \rho{^I}_{d}}{\delta \rho{^{N}}_{a}}
\frac{\delta \rho{^{II}}_{e}}{\delta \rho{^{N}}_{b}}\nonumber\\
&=& 
\Omega{^{N}}{_{a_1\dots a_Ma_{M+1}\dots a_N,\,
b_1\dots b_{M+K}b_{M+K+1}\dots b_N,\,
c_1\dots c_N
}}
\delta^{a_1}_{d_1}\dots \delta^{a_M}_{d_M}
\omega^{a_{M+1}\dots a_N}
\omega^{b_{1}\dots b_{M+K}}
\delta^{b_{M+K+1}}_{e_{M+K+1}}\dots \delta^{b_N}_{e_N}\nonumber\\
&=& 
\Omega{^{N}}{_{d_1\dots d_Ma_{M+1}\dots a_N,\,
b_1\dots b_{M+K}e_{M+K+1}\dots e_N,\,
c_1\dots c_N
}}
\omega^{a_{M+1}\dots a_N}
\omega^{b_{1}\dots b_{M+K}}\nonumber\\
&=& 
\Bigl(
I_{\delta_1\beta'_1}\dots I_{\delta_M\beta'_M}
I_{\alpha_{M+1}\beta'_{M+1}}\dots I_{\alpha_{M+K}\beta'_{M+K}}
I_{\alpha_{M+K+1}\ve'_{M+K+1}}\dots I_{\alpha_{N}\beta'_{N}}
\nonumber\\
&\pp =&\pp {xxxxxx}\times
I_{\beta_1\gamma'_1}\dots I_{\beta_{M+K}\gamma'_{M+K}}
I_{\ve_{M+K+1}\gamma'_{M+K+1}}\dots I_{\ve_N\gamma'_N}
\nonumber\\
&\pp =&\pp {xxxxxxxxxxxxxxxxxxxxxxx}\times
I_{\gamma_1\delta'_1}\dots I_{\gamma_{M}\delta'_M}
I_{\gamma_{M+1}\alpha'_{M+1}}\dots I_{\gamma_N\alpha'_N}\nonumber\\
&\pp =& -
I_{\delta_1\gamma'_1}\dots I_{\delta_{M}\gamma'_M}
I_{\alpha_{M+1}\gamma'_{M+1}}\dots I_{\alpha_N\gamma'_N}
\nonumber\\
&\pp =&\pp {xxxxxx}\times
I_{\beta_1\delta'_1}\dots I_{\beta_M\delta'_M}
I_{\beta_{M+1}\alpha'_{M+1}}\dots I_{\beta_{M+K}\alpha'_{M+K}}
I_{\ve_{M+K+1}\alpha'_{M+K+1}}\dots I_{\beta_{N}\alpha'_{N}}
\nonumber\\
&\pp =&\pp {xxxxxxxxxxxxxxxxxxxxxxx}\times
I_{\gamma_1\beta'_1}\dots I_{\gamma_{M+K}\beta'_{M+K}}
I_{\gamma_{M+K+1}\ve'_{M+K+1}}\dots I_{\gamma_N\ve'_N}
\Bigr)
\nonumber\\
&\pp =&\pp {xxxxxxxxxxxxxx}\times
\omega^{\alpha_{M+1}\alpha'_{M+1}}\dots 
\omega^{\alpha_{N}\alpha'_{N}}
\omega^{\beta_{1}\beta'_{1}}\dots 
\omega^{\beta_{M+K}\beta'_{M+K}}\nonumber\\
&=&
I_{\delta_1\beta'_1}\dots I_{\delta_M\beta'_M}
\delta^{\alpha'_{M+1}}_{\beta'_{M+1}}\dots 
\delta^{\alpha'_{M+K}}_{\beta'_{M+K}}
\delta^{\alpha'_{M+K+1}}_{\ve'_{M+K+1}}\dots 
\delta^{\alpha'_{N}}_{\beta'_{N}}
\nonumber\\
&\pp =&\pp {xxxxxx}\times
\delta^{\beta'_1}_{\gamma'_1}\dots 
\delta^{\beta'_{M+K}}_{\gamma'_{M+K}}
I_{\ve_{M+K+1}\gamma'_{M+K+1}}\dots I_{\ve_N\gamma'_N}
\nonumber\\
&\pp =&\pp {xxxxxxxxxxxxxxxxxxxxxxx}\times
I_{\gamma_1\delta'_1}\dots I_{\gamma_{M}\delta'_M}
I_{\gamma_{M+1}\alpha'_{M+1}}\dots I_{\gamma_N\alpha'_N}\nonumber\\
&\pp =& -
I_{\delta_1\gamma'_1}\dots I_{\delta_{M}\gamma'_M}
\delta^{\alpha'_{M+1}}_{\gamma'_{M+1}}\dots 
\delta^{\alpha'_N}_{\gamma'_N}
\nonumber\\
&\pp =&\pp {xxxxxx}\times
\delta^{\beta'_1}_{\delta'_1}\dots 
\delta^{\beta'_M}_{\delta'_M}
\delta^{\beta'_{M+1}}_{\alpha'_{M+1}}\dots 
\delta^{\beta'_{M+K}}_{\alpha'_{M+K}}
I_{\ve_{M+K+1}\alpha'_{M+K+1}}\dots I_{\beta_{N}\alpha'_{N}}
\nonumber\\
&\pp =&\pp {xxxxxxxxxxxxxxxxxxxxxxx}\times
I_{\gamma_1\beta'_1}\dots I_{\gamma_{M+K}\beta'_{M+K}}
I_{\gamma_{M+K+1}\ve'_{M+K+1}}\dots I_{\gamma_N\ve'_N}
=0.
\ee
The proof is completed by
$
[\rho{^{I}}_{d},\rho{^{II}}_{e},\,\cdot\,]^N=X_c
\delta/\delta\rho_c=0.
$
$\Box$

\medskip
A straightforward consequence of Lemma~1 is the following
important theorem about nonexistence of ``faster-than-light
telegraphs" for all Nambu-type generalizations of the Dirac
equation. 

\medskip\noindent
{\bf Theorem 2.} Consider two, in general nonlinear, observables 
$F^{I}[\rho]=F^{I}[\rho^{I}]$, $G^{II}[\rho]=G^{II}[\rho^{II}]$
corresponding 
to two nonoverlapping, $M$- and $(N-M-K)$-particle subsystems
of a larger $N$-particle system. Then
\be
[F^{I},G^{II},\,\cdot\,]^N=0.
\ee
{\it Proof\/}: 
\be
[F^{I},G^{II},\,\cdot\,]^N=
 \Omega{^{N}}{_{abc}}
\frac{\delta \rho{^I}_{d}}{\delta \rho{^{N}}_{a}}
\frac{\delta \rho{^{II}}_{e}}{\delta \rho{^{N}}_{b}}
\frac{\delta F^{I}}{\delta \rho{^I}_{d}}
\frac{\delta G^{II}}{\delta \rho{^{II}}_{e}}
\frac{\delta}{\delta\rho_c}=0.
\ee
$\Box$

\medskip\noindent
The meaning of Theorem~2 is the following. Consider two
noninteracting subsystems described by a 
(possibly nonlinear) Hamiltonian function 
\be
H[\rho]=H^{I}[\rho^{I}] + H^{II}[\rho^{II}].
\ee
Then, for {\it any\/} $S$
\be
i\dot F^{I}=[F^{I}, H, S]=[F^{I}, H^{I}, S]
\ee
and the dynamics of a subsystem is generated by the
Hamiltonian fuction of this subsystem. Theorem~2 is a
generalization of the theorems of Polchinski \cite{Polchinski},
which was demonstrated for pure-state density matrices,
and Jordan \cite{Jordan} which was formulated in terms of
arbitrary density matrices. The results of Polchinski and Jordan
referred to Weinberg's nonlinear quantum mechanics, which is a
particular case of our triple-bracket formulation obtained if
we put $S=C_2/2$ (in our case $S$ is arbitrary). 
In addition all those formulations were two-particle and
nonrelativistic. Our theorem is the first step towards Fock space
and, in larger perspective, 
relativistic field-theoretic generalization. 
This result is also interesting in the context of 
Weinberg's remark that he ``could not find any way to extend the
nonlinear version of quantum mechanics to theories based on
Einstein's special theory of relativity" \cite{Dreams}. 

\medskip\noindent
{\bf Lemma 3.}
\be
[C_n,C_m,\,\cdot\,]^N=0.
\ee
{\it Proof\/}: To simplify notation we shall not explicitly
write the number-of-particles index $N$ in formulas. 
The indices $a,\,b,\,c,a_k,\,b_l$ are themselves $N$-particle indices.
We have
\be
\frac{\delta C_n}{\delta \rho_{a}}=
ng^{b_1\dots b_{n-1}a}\rho_{b_1}\dots \rho_{b_{n-1}}.
\ee
Consider the expression
\be
Y^{m,n}_c&=&\Omega_{abc}g^{a_1\dots a_na}g^{b_1\dots b_mb}
\rho_{a_1}\dots \rho_{a_n}
\rho_{b_1}\dots \rho_{b_m}\nonumber\\
&=&
\Bigl(
I_{\alpha\beta'}I_{\beta\gamma'}I_{\gamma\alpha'}
-
I_{\alpha\gamma'}I_{\beta\alpha'}I_{\gamma\beta'}
\Bigr)
\omega^{\alpha_1\alpha'}\omega^{\alpha_2\alpha'_1}
\dots
\omega^{\alpha_n\alpha'_{n-1}}\omega^{\alpha\alpha'_n}
\omega^{\beta_1\beta'}\omega^{\beta_2\beta'_1}
\dots
\omega^{\beta_m\beta'_{m-1}}\omega^{\beta\beta'_m}
\nonumber\\
&\pp =&\pp {xxxxxxxxx}\times 
\rho_{a_1}\dots \rho_{a_n}
\rho_{b_1}\dots \rho_{b_m}
\nonumber\\
&=&
\delta^{\alpha_1}_{\gamma}
\omega^{\alpha_2\alpha'_1}\omega^{\alpha_3\alpha'_2}
\dots
\omega^{\alpha_n\alpha'_{n-1}}
\omega^{\beta_1\alpha'_n}\omega^{\beta_2\beta'_1}
\dots
\omega^{\beta_m\beta'_{m-1}}
\delta^{\beta'_m}_{\gamma'}
\rho_{\alpha_1\alpha'_1}\dots \rho_{\alpha_n\alpha'_n}
\rho_{\beta_1\beta'_1}\dots \rho_{\beta_m\beta'_m}\nonumber\\
&\pp =&
-
\delta^{\beta_1}_{\gamma}
\omega^{\beta_2\beta'_1}\omega^{\beta_3\beta'_2}
\dots
\omega^{\beta_m\beta'_{m-1}}
\omega^{\alpha_1\beta'_m}\omega^{\alpha_2\alpha'_1}
\dots
\omega^{\alpha_n\alpha'_{n-1}}
\delta^{\alpha'_m}_{\gamma'}
\rho_{\alpha_1\alpha'_1}\dots \rho_{\alpha_n\alpha'_n}
\rho_{\beta_1\beta'_1}\dots \rho_{\beta_m\beta'_m}
\nonumber\\
&=&
\omega^{\alpha_2\alpha'_1}
\dots
\omega^{\alpha_n\alpha'_{n-1}}
\omega^{\alpha_{n+1}\alpha'_n}\omega^{\alpha_{n+2}\alpha'_{n+1}}
\dots
\omega^{\alpha_{n+m}\alpha'_{n+m-1}}
\rho_{\gamma\alpha'_1}\dots \rho_{\alpha_{n+m}\gamma'}
\label{linia1}\\
&\pp =&\pp {xxxx} -
\omega^{\beta_2\beta'_1}
\dots
\omega^{\beta_m\beta'_{m-1}}
\omega^{\beta_{m+1}\beta'_m}\omega^{\beta_{m+2}\beta'_{m+1}}
\dots
\omega^{\beta_{n+m}\beta'_{n+m-1}}
\rho_{\gamma\beta'_1}\dots \rho_{\beta_{n+m}\gamma'}=0.
\label{linia2}
\ee
where we have renamed indices:
 $(\beta_1,\dots \beta_m)\to (\alpha_{n+1},\dots
\alpha_{n+m})$ in (\ref{linia1}), and 
$(\alpha_1,\dots \alpha_n)\to (\beta_{m+1},\dots
\beta_{n+m})$ in (\ref{linia2}). Now
\be
[C_m,C_n,\,\cdot\,]^N=mnY^{m-1,n-1}_c\frac{\delta}{\delta
\rho_c}=0 \nonumber
\ee
which completes the proof.$\Box$

The consequence of Lemma~3 is

\medskip\noindent
{\bf Theorem~4.} Let $S=S(C_1,\dots C_k,\dots)$ be any
differentiable function of $C_1,\dots $. Then
\be
[C_n,\,\cdot\,,S]^N=0.
\ee
{\it Proof\/}: 
\be
[C_n,\,\cdot\,,S]^N=\sum_k[C_n,\,\cdot\,,C_k]^N
\frac{\partial S}{\partial C_k}=0.\nonumber
\ee
$\Box$

As a consequence, $C_n$ are constants of motion for all
(in general nonlinear) Hamiltonian functions $H$ if the generalized
entropies depend on $\rho_a$ via $C_k$. $C_n$ are, for such
a class of entropies, 
``Casimir
invariants" of the triple bracket algebra of observables. 
These results generalize the theorem of Jordan
\cite{Jordan} who proved this property of $C_n$ for $S=C_2/2$
in nonrelativistic nonlinear quantum mechanics. 

\section{Density matrix interpretation of solutions of the
generalized Dirac equation}
\label{s4}

The fact that $C_n=\Tr (\rho^n)$ are constants of motion for any
$H$ if $S$ depends on $\rho_a$ via $C_k$, $k=1,2,\dots$, can
be used to prove the following

\medskip\noindent
{\bf Theorem~5.} Let $t\mapsto \rho_a(t)$ be a Hermitian
solution of
\be
i \dot \rho_a = [\rho_a,H,S]
\ee
where $H$ is arbitrary and $S=S(C_1,\dots C_k,\dots)$, such that
$\rho_a(0)$ is a density matrix having a finite number of
nonvanishing eigenvalues $p_k(0)$. Then eigenvalues $p_k(t)$
of $\rho_a(t)$ are constants of motion, i.e. $p_k(t)=p_k(0)$ for
$t>0$.

\noindent{\it Proof\/}: Since the nonvanishing eigenvalues of $\rho_0$
satisfy $0<p_k(0)\leq 1<2$, it follows that for any $\alpha$
$p_k(0)^\alpha$ can be written in the form of a convergent Taylor
series. By virtue of the spectral theorem the same is true for
$\rho_0^\alpha$ and ${\rm Tr}\, (\rho_0^\alpha)$. Each term
of the Taylor expansion of ${\rm Tr}\, (\rho_0^\alpha)$ is
proportional to $f_n[\rho_0]$, for some $n$. But
$f_n[\rho_0]=f_n[\rho_t]$ hence
\begin{equation}
{\rm Tr}\, (\rho_0^\alpha)={\rm Tr}\, (\rho_t^\alpha)=\sum_k
p_k(0)^\alpha=
\sum_k p_k(t)^\alpha
\end{equation}
for all real $\alpha$. Since all $p_k(0)$ are assumed to be
known (the initial 
condition), we know also $\sum_k p_k(0)^\alpha=\sum_k
p_k(t)^\alpha$ for any $\alpha$. We can now apply the  result
from information theory \cite{Renyi}  that the
knowledge of 
\begin{equation}
\sum_{k=1}^{n<\infty} p_k(t)^\alpha\label{n<}
\end{equation}
 for all $\alpha$ {\it
uniquely\/} determines $p_k(t)$. The continuity in $t$ implies
that $p_k(t)=p_k(0)$.$\Box$

{\it Remark\/}: The assumption that initially the 
density matrix has a {\it finite\/}
number of nonvanishing 
eigenvalues $p_k(0)$ is necessary since the 
theorem we use in the proof is formulated in \cite{Renyi}
 for sums  (\ref{n<}) with 
finite $n$. This result is not fully satisfactory, but is
sufficient at least ``for all practical purposes". 

\section{Two examples}
\label{s5}

Consider now the generalization where observables are linear 
and, in addition, a multiplication of a solution 
by a number is a symmetry of the evolution equation. To maintain
the latter a homogeneity of $S$ must be the same as this of
$C_2$. 
An example of 
 homogeneity preserving generalization of $S=C_2/2$ 
is
\begin{equation}
S_\alpha[\rho]=\Bigl(1-\frac{1}{\alpha}\Bigr)
\frac{\bigl({\rm Tr}\,(\rho^\alpha)\bigr)^{1/(\alpha-1)}}{({\rm
Tr}\,\rho) 
^{1/(\alpha-1)-1}}\label{S-alpha}
\end{equation}
where $\rho$ is a density matrix.
The choice of the denominator is important only from the point
of view of the homogeneity of the evolution equation. The
multiplier $1-1/\alpha$ guarantees that the evolution of pure
states is the same, and therefore {\it linear\/}, for all $\alpha$. 
The generalized Liouville-von Neumann
equation following from (\ref{S-alpha}) is
\begin{equation}
i\dot \rho=\frac{\bigl({\rm
Tr}\,(\rho^\alpha)\bigr)^{1/(\alpha-1)-1}} {({\rm Tr}\,\rho)
^{1/(\alpha-1)-1}}[\hat H,\rho^{\alpha-1}].
\end{equation}
For pure states and ${\rm Tr}\,\rho=1$, $\rho^n=\rho$ and the
equation reduces to the ordinary, linear one; for mixed states
the evolution is nonlinear unless the states are 
mixed to the extent
that $\rho$ is proportional to the unit operator.

The evolution of the (linear) observables is governed by
\begin{equation}
i\dot F=\frac{\bigl({\rm
Tr}\,(\rho^\alpha)\bigr)^{1/(\alpha-1)-1}} {({\rm Tr}\,\rho)
^{1/(\alpha-1)-1}}{\rm Tr}\,\bigl(\rho^{\alpha-1}[\hat F,\hat
H]\bigr)
\end{equation}
which shows that for the generalized $S$ the time derivative of
an observable is not linear in the density matrix.  For
$\alpha=2$ the equations reduce  to the ordinary linear
equations.

Also of interest is the following choice of $S_\alpha$
\begin{equation}
S_\alpha[\rho]={1\over 2}
\frac{\bigl({\rm Tr}\,(\rho^\alpha)
\bigr)^{1/(\alpha-1)}}{({\rm Tr}\,\rho)
^{1/(\alpha-1)-1}}.
\end{equation}
For pure states the expression reduces to the linear form
${1\over 2}
\langle\psi|\psi\rangle^2={1\over 2}{\rm Tr}\,(\rho^2)$, 
and the density matrix then satisfies
\begin{equation}
i\dot \rho= {1\over 2}\frac{\alpha}{\alpha-1}
\frac{\bigl({\rm Tr}\,(\rho^\alpha)\bigr)^{1/(\alpha-1)-1}}
{({\rm Tr}\,\rho) ^{1/(\alpha-1)-1}}[\hat
H,\rho^{\alpha-1}]\label{S-alpha'}
\end{equation}
which for pure states and normalized $\rho$  becomes
\begin{equation}
2\frac{\alpha -1}{\alpha}i\dot \rho= [\hat H,\rho].
\end{equation}

\acknowledgments
I am grateful to my advisor Prof. Kazimierz Rz\c a\.zewski,
Prof. Iwo Bia{\l}ynicki-Birula, and Nicolas Gisin for
their valuable comments and help, and to Drs. E. Garcia Alvarez
and F. Gaioli for comments on the Hamiltonian formulation of the
Dirac equation. 
I am deeply indebted to my MIT friends:
 Mike Bradley, Mike Chapman, Troy Hammond,
Al Lenef, Fred Palmer, and Ed Smith for reading an earlier
version of the paper. 
Special thanks  to Mike Andrews for his constant 
help and patience. Finally, 
I would like to thank Prof.~David~E.~Pritchard for his 
hospitality at MIT, where the paper was completed, and
the Fulbright Commission for funding.

\section{Appendix: Infeld-van der Waerden tensors and generators of
(1/2,0) and (0,1/2)} \label{A1}

This Appendix explains spinor conventions used in this Letter
Consider  representations $(\frac{1}{2},0)$ and 
 $(0,\frac{1}{2})$
 of an element $\omega\in SL(2,C)$: $e^{\frac{i}{2}
\omega^{ab}{\sigma}_{ab}}$ and 
$e^{\frac{i}{2}
 \omega^{ab}\bar {\sigma}_{ab}}$.
 The explicit form of the generators in terms of 
Infeld-van der Waerden tensors is
\begin{eqnarray}
\frac{1}{2i}\Bigl(g^a_{\pp A XA'}g^{bYA'}-g^b_{\pp A
XA'}g^{aYA'}\Bigr)
&=&{ \sigma}^{ab\pp A
Y}_{\pp {aa}X},\\
\frac{1}{2i}\Bigl(
g^a_{\pp A AX'}g^{bAY'}-g^b_{\pp A AX'}g^{aAY'}\Bigr)&=&\bar
{ \sigma}^{ab\pp A
Y'}_{\pp {aa}X'}.
\end{eqnarray}
Their purely spinor form is
\begin{eqnarray}
{\sigma}_{AA'BB'XY}&=&\frac{1}{2i}\varepsilon
_{A'B'}\bigl(\varepsilon_{AX}\varepsilon_{BY}+
\varepsilon_{BX}\varepsilon_{AY}\bigr),\label{gs1}\\
\bar {\sigma}_{AA'BB'X'Y'}
&=&\frac{1}{2i}\varepsilon_{AB}\bigl(\varepsilon_{A'X'}\varepsilon_{B'Y'}+
\varepsilon_{B'X'}\varepsilon_{A'Y'}\bigr)\label{gs2},
\end{eqnarray}
Dual tensors are $^*\bar
{\sigma}^{ab\pp A
Y'}_{\pp {aa}X'}=+i\bar
{\sigma}^{ab\pp A
Y'}_{\pp {aa}X'}$
and 
$^*{\sigma}^{ab\pp A
Y}_{\pp {aa}X}=-i{\sigma}^{ab\pp A
Y}_{\pp {aa}X}$.

Additionally the Infeld-van der Waerden tensors satisfy
\begin{eqnarray}
g^a_{\pp A XA'}g^{bYA'}+g^b_{\pp A
XA'}g^{aYA'}
&=&g^{ab}\varepsilon^{{\pp {X}}
Y}_{X}\label{IW1}\\
g^a_{\pp A AX'}g^{bAY'}+g^b_{\pp A
AX'}g^{aAY'}
&=&g^{ab}\varepsilon^{{\pp {X}}
Y'}_{X'}\label{IW2}
\end{eqnarray}
These equations lead to the identities
\be
g^a_{\pp A XA'}g^{bYA'}&=&\frac{1}{2}g^{ab}\varepsilon^{{\pp {X}}
Y}_{X}+i{ \sigma}^{ab\pp A
Y}_{\pp {aa}X}\label{id1}\\
g^a_{\pp A AX'}g^{bAY'}&=&\frac{1}{2}g^{ab}\varepsilon^{{\pp {X}}
Y'}_{X'}+i
\bar
{ \sigma}^{ab\pp A
Y'}_{\pp {aa}X'}\label{id2}
\ee

\end{document}